\documentclass[3p,times]{elsarticle}
\usepackage{lineno,hyperref,multirow,graphicx,booktabs }
\modulolinenumbers[5]

\journal{Physica A: Statistical Mechanics and its Applications}

\begin{document}
\begin{frontmatter}
\title{A vertex similarity index for better personalized recommendation}

\author[inst1,inst2]{Ling-Jiao Chen}
\author[inst3]{Zi-Ke Zhang}
\author[inst1,inst2]{Jin-Hu Liu}
\author[inst1,inst2]{Jian Gao\corref{cor1}}
\cortext[cor1]{Corresponding author. \\ \hspace*{0.4cm} \emph{E-mail address}: gaojian08@hotmail.com (J. Gao)}
\author[inst1,inst2]{Tao Zhou}

\address[inst1]{CompleX Lab, Web Sciences Center, University of Electronic Science and Technology of China, Chengdu 611731, People's Republic of China}
\address[inst2]{Big Data Research Center, University of Electronic Science and Technology of China, Chengdu 611731, People's Republic of China}
\address[inst3]{Institute of Information Economy, Hangzhou Normal University, Hangzhou 310036, People's Republic of China}

\begin{abstract}
Recommender systems benefit us in tackling the problem of information overload by predicting our potential choices among diverse niche objects. So far, a variety of personalized recommendation algorithms have been proposed and most of them are based on similarities, such as collaborative filtering and mass diffusion. Here, we propose a novel vertex similarity index named CosRA, which combines advantages of both the cosine index and the resource-allocation (RA) index. By applying the CosRA index to real recommender systems including MovieLens, Netflix and RYM, we show that the CosRA-based method has better performance in accuracy, diversity and novelty than some benchmark methods. Moreover, the CosRA index is free of parameters, which is a significant advantage in real applications. Further experiments show that the introduction of two turnable parameters cannot remarkably improve the overall performance of the CosRA index.
\end{abstract}


\end{frontmatter}
\linenumbers
\nolinenumbers

\section{Introduction}
The development of the Internet and e-commerce makes our lives more convenient as billions of products are available online \cite{Laudon2007}. Meanwhile, the problem of information overload plagues us everyday as it is much harder to dig out relevant objects than ever \cite{Zhang2013}. Thus far, personalized recommendation was thought to be the most promising way to efficiently solve the problem of information overload \cite{Schafer1999,Porcel2010}. Personalized recommendation benefits both buyers and sellers, and it is now playing an increasing role in our online social lives. Many online platforms (Amazon, eBay, AdaptiveInfo, Taobao, etc) have introduced personalized recommendation systems \cite{Kantor2011}, which predict users' potential choices by analyzing historical behaviors of users, attributes of objects, and so on \cite{Gualdi2013}. For example, Amazon.com recommends books by analyzing users' purchase records \cite{Linden2003}, and AdaptiveInfo.com recommends news by using users' reading histories \cite{Billsus2002}. In recent years, personalized recommendation has found wide applications \cite{Schafer2001} in recommending movies \cite{Liu2013,LiuJ2014}, videos \cite{Davidson2010}, research articles \cite{Bogers2008}, driving routes \cite{Ge2010}, locations \cite{Lian2015c,Lian2015} and so on.

So far, a variety of personalized recommendation algorithms have been proposed \cite{Lu2012,Bobadilla2013,Adomavicius2005,Zeng2014,Gan2014}, among which user-based (UCF) and item-based collaborative filtering (ICF) are the most representative ones \cite{Sarwar2001}. UCF and ICF are respectively based on the weighted combination of similar users' opinions and the similarity between items \cite{Goldberg1992}. Recently, many diffusion-based algorithms are proposed by introducing some physical dynamics into the recommender systems, such as mass diffusion (MD) \cite{Zhang2007} and heat conduction (HC) \cite{Zhang2007h}. The simplest version of MD can be considered as a two-step resource-allocation process in bipartite networks \cite{Zhou2007}. Later, Zhou \emph{et al.} \cite{Zhou2008} and Jia \emph{et al.} \cite{Jia2008} proposed two algorithms by giving new strategies in the initial resource distribution, Zhou \emph{et al.} \cite{ZhouT2010} proposed a hybrid method that combines both MD and HC, L{\"u} \emph{et al.} \cite{Lu2011} proposed a preferential diffusion method by considering node weights in redistributing resources, and Liu \emph{et al.} \cite{Liu2011b} proposed a weighted heat conduction algorithm by considering edge weighting. Reviews of previous literatures can be found in Refs. \cite{Lu2012} and \cite{Bobadilla2013}.

Essentially, the aforementioned collaborative filtering and diffusion-based methods are based on similarities \cite{Deshpande2004,Guo2014}. In collaborative filtering, the most commonly used index is cosine similarity \cite{Ahn2008,Choi2013,Bagchi2015}. However, it strongly tends to recommend popular objects, resulting in accurate yet less-diverse recommendations \cite{Pan2010}. In diffusion-based methods, the diffusion is indeed a resource-allocation process, and the node similarity is characterized by the resource-allocation (RA) index \cite{Ou2007,Zhou2009}. The RA index gives high priority to assign resources to large-degree nodes, which leads to high accuracy but low diversity of MD \cite{Liu2011}. In fact, the cosine index and RA index are complementary to each other, and thus to combine the two can possibly improve the overall performance. How to design a suitable similarity index for better recommendation is still an open issue and such index can be applied in characterizing many network structures and functions \cite{Scholz2010,Leicht2006}.

In this paper, we propose a vertex similarity index, named CosRA, for better personalized recommendation. Based on the CosRA index which combines advantages of both the cosine index and the RA index, we further propose a personalized recommendation algorithm. Extensive experiments on four real data sets suggest that the CosRA-based method performs better in accuracy, diversity and novelty than some benchmark methods. Moreover, we provide some insights on the mechanism of the CosRA index and extend it to a more general form by introducing two turnable parameters. Interestingly, results suggest that the original CosRA index is almost optimal, and its effectiveness cannot be remarkably improved by adjusting the parameters. Such feature is significant since a parameter-free index is more applicable than a parameter-dependent index. Our work sheds lights on the importance of a suitable vertex similarity index in enhancing the overall performance of personalized recommendation.

\section{Vertex similarity index}
A recommender system can be naturally described by a user-object bipartite network $G(U, O, E)$, where $U=\{u_{1}, u_{2}, \ldots, u_{m}\}$, $O=\{o_{1}, o_{2}, \ldots, o_{n}\}$ and $E=\{e_1, e_2, \ldots, e_z\}$ are sets of users, objects and links, respectively. To distinguish object-related and user-related indices, we respectively use Greek and Latin letters for them. Meanwhile, the bipartite network $G(U, O, E)$ can be naturally represented by an adjacency matrix $A$, whose element $a_{i\alpha}=1$ if there is a link connecting node $U_{i}$ and node $O_{\alpha}$, \emph{i.e.}, user $i$ has collected object $\alpha$, otherwise $a_{i\alpha}=0$. The main purpose of recommendation algorithms is to provide a target user with a ranking list of his uncollected objects. For user $i$, the recommendation list with length $L$ is denoted as $o_i^L$. That is to say, $o_i^L$ is a set of $L$ objects with the highest recommendation scores for user $i$.

First, we introduce two widely used similarity indices in recommendation algorithms, namely, the cosine index and the RA index. Taking two objects $\alpha$ and $\beta$ as an example, the cosine index between them is defined as
\begin{equation}
S_{\alpha\beta}^{Cos} = \frac{1}{\sqrt{k_{\alpha} k_{\beta}}} \sum_{i=1}^{m} a_{i\alpha} a_{i\beta},
\label{eq:cos}
\end{equation}
where $k_{\alpha}$ and $k_{\beta}$ are the degrees of objects $\alpha$ and $\beta$, respectively. In fact, the cosine index measures the similarity between two objects' rating vectors of an inner product space. Meanwhile, the resource-allocation process is equivalent to the one-step random walk in the user-object bipartite networks starting from the common neighbors \cite{Zhou2009}. Specifically, the RA index between two objects $\alpha$ and $\beta$ is defined as
\begin{equation}
S_{\alpha\beta}^{RA} = \sum_{i=1}^{m} \frac{a_{i\alpha} a_{i\beta}}{k_i},
\label{eq:ra}
\end{equation}
where $k_i$ is the degree of user $i$. Indeed, the RA index is the entry of the transformation matrix in the simplest version of the MD process \cite{Zhou2007}, which is a variant on an earlier version of the probabilistic spreading algorithm \cite{ZhouT2010}.

Then, we introduce the proposed CosRA similarity index. On the one side, both of the degrees of the two objects should be considered, and the effect of popular objects should be restricted in calculating the similarity. On the other side, the effect of small-degree users should be enhanced to decrease the advantage of large-degree nodes in the network. Based on these two considerations, the CosRA index is proposed by combining both the cosine index and the RA index. Specifically, for two objects $\alpha$ and $\beta$, the CosRA index is defined as
\begin{equation}
S_{\alpha\beta}^{CosRA} = \frac{1}{\sqrt{k_{\alpha} k_{\beta}}} \sum_{i=1}^{m} \frac{a_{i\alpha} a_{i\beta}}{k_i}.
\label{eq:cosra}
\end{equation}
where $k_i$ is the degree of user $i$, $k_{\alpha}$ is the degree of object $\alpha$, and $k_{\beta}$ is the degree of object $\beta$. Actually, $S_{\alpha\beta}^{CosRA}$ measures the similarity between objects $\alpha$ and $\beta$ by summing their contribution from all two-step paths considering the degrees of both types of nodes in bipartite networks.

\begin{figure}
\centering
\includegraphics[width=0.55\textwidth]{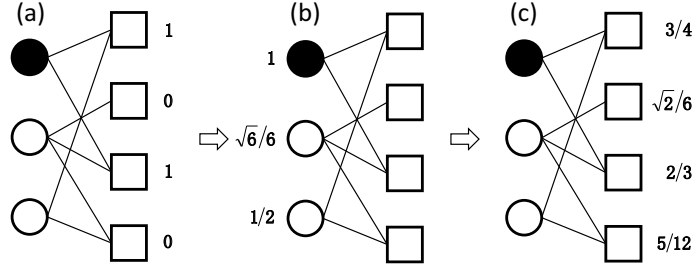}\\
\caption{Illustration of the proposed CosRA-based method. (a) Initially, for a target user (colored black), the resource of each object is initialized by Eq.~(\ref{eq:init}). (b) Then, objects distribute the resources to all the users who have collected them. (c) Finally, all users redistribute the resources to the objects that they have collected. The processes in panels (b) and (c) can be characterized by Eq.~(\ref{eq:tf}).}
\label{fig:EW}
\end{figure}

Further, we propose a personalized recommendation algorithm based on the CosRA index. Specifically, the proposed CosRA-based method works as follows: Firstly, for user $i$, the resource of object $\alpha$ is initialized as
\begin{equation}
f_{\alpha}^{(i)}= a_{i\alpha},
\label{eq:init}
\end{equation}
where $a_{i\alpha}=1$ if user $i$ has collected object $\alpha$, otherwise $a_{i\alpha}=0$. Secondly, the resources of all objects are redistributed via the transformation
\begin{equation}
f'^{(i)}= S^{CosRA} f^{(i)},
\label{eq:tf}
\end{equation}
where $f^{(i)}$ is an $n$-dimensional vector recording all objects' initial resources given the target user $i$, and $f'^{(i)}$ is the vector recording all the final resources that located on each object. Finally, all objects are sorted by their final resources $f'^{(i)}$, and then the top-$L$ uncollected objects are recommended to user $i$. An illustration of the CosRA-based method is shown in Fig.~\ref{fig:EW}.

\section{Data and evaluation}
\subsection{Data descriptions}
Four commonly studied real data sets, namely, MovieLens-\scriptsize{100K}\normalsize{, MovieLens-}\scriptsize{1M}\normalsize{,} Netflix and RYM, are used to test the performance of different methods. MovieLens data set is provided by the GroupLens project at University of Minnesota\footnote{http://www.grouplens.org.}. The data set uses a 5-point rating scale from 1 to 5 (\emph{i.e.}, worst to best). When building bipartite networks, we only consider the links with ratings $\geq3$. After coarse graining, the smaller data set contains 82520 links and the larger one contains 836478 links. Netflix data set is released by the DVD rental company Netflix for its Netflix Prize contest in \emph{Netflix.com}\footnote{http://www.netflixprize.com.}. The ratings are also given on a 5-point scale. Analogously, only links with ratings $\geq3$ are reserved. Then, we extract a small sampling data set by randomly choosing 10000 users and taking the associated 701947 links. RYM data set is publicly available on the music ratings website \emph{RateYourMusic.com}\footnote{http://www.rateyourmusic.com.}. The ratings are given on a 10-point scale from 1 to 10 (\emph{i.e.}, worst to best). Here, only links with ratings $\geq6$ are considered, and thus the final data contains 609792 links. Note that, the bipartite networks in the following analysis are unweighted with rating values being neglected. The basic statistics of the data sets are summarized in Table~\ref{tab1:data}.

\begin{table}
    \caption{Basic statistics of the four real online rating data sets.}
    \begin{tabular*}{\textwidth}{@{\extracolsep{\fill}}lllll}
    \toprule
    Data & Users & Objects & Links & Sparsity \\
    \midrule
    MovieLens-\scriptsize{100K} & 943 & 1574 & 82520 & $5.56\times10^{-2}$\\
    MovieLens-\scriptsize{1M} & 6039 & 3628 & 836478 & $3.82\times10^{-2}$\\
    Netflix & 10000 & 5640 & 701947 & $1.24\times10^{-2}$\\
    RYM & 33197 & 5234 & 609792 & $3.51\times10^{-3}$\\
    \bottomrule
    \end{tabular*}
\label{tab1:data}
\end{table}

\subsection{Metrics for evaluation}
In order to estimate how accurately the recommendation algorithms will perform in practice, the cross-validation is usually used for assessing how the results will generalize to an independent data set \cite{McLachlan2005}. One round of cross-validation involves partitioning a sample of data into complementary subsets, performing the analysis on one subset (namely, training set), and validating the analysis on the other subset (namely, testing set). In the following experiments, we use a 10-folder cross-validation strategy to evaluate the performance of recommendation in each independent realization. Specifically, all the ratings of users (i.e., links in bipartite networks) are randomly split into 10 equal sized subsamples, independent of users and objects. Then, of the 10 subsamples, one subsample is retained for testing the recommendation performance, and the remaining 9 subsamples are combined and used as the training set for recommendation. In another word, 90\% of the whole data sets are used for recommendation and the other 10\% are used for evaluation. In this way, the 10-folder cross-validation process is repeated 10 times, with each of the 10 subsamples used exactly once as testing data. Finally, all the 10 results are averaged to produce one single result for this realization.

Evaluation metrics about the recommendation performance have been widely investigated in previous literatures \cite{Zhou2007,Liu2010}. In this paper, we apply seven widely used metrics to quantify the performance of recommender systems, including four accuracy metrics (AUC, MAP, Precision and Recall), two diversity metrics (Inter-similarity and Intra-similarity), and one novelty metric (Popularity). In the following, we will briefly introduce these metrics.

Accuracy is one of the most important metric in evaluating the quality of recommendation algorithms. We first introduce AUC (area under the ROC curve) \cite{Hanley1982}. Given the ranks of objects in the testing set, AUC value can be interpreted as the probability that a randomly chosen collected object is ranked higher than a randomly chosen un-collected object. To calculate AUC, at each time, a pair of collected and un-collected objects is selected to compare their resources. After $N$ times independent comparisons, if there are $N_1$ times the collected object has more resources and $N_2$ times their resources are the same, the average value of AUC for all users is defined as \cite{Zhou2009}
\begin{equation}
AUC = \frac{1}{m}\sum_{i=1}^{m}\frac{(N_1 + 0.5N_2)}{N}.
\label{eq.6}
\end{equation}
Larger AUC value means higher algorithmic accuracy.

Then we introduce three $L$-dependent accuracy metrics, namely, MAP (Mean Average Precision) \cite{Liu2007}, Precision and Recall \cite{Herlocker2004}. MAP is a standard rank-aware measure of the overall ranking accuracy in the field of information retrieval, which is similar to the average ranking score \cite{Zhou2007,Liu2010}. The average precision for user $i$ is defined as
\begin{equation}
\bar{P}_i(L) = \frac{1}{D(i)}\sum_{s=1}^{d_{i}(L)}\frac{s}{r_s},
\label{eq.7.1}
\end{equation}
where $D(i)$ is the number of objects in the testing set, $d_i(L)$ is the number of common objects in the testing set and the recommendation list with length $L$, and $r_{s}\in [1,L]$ is the rank of $s$-th common object in the recommendation list. Then, the MAP index is calculated by averaging $\bar{P}_i(L)$ for all users via
\begin{equation}
\mbox{MAP}=\frac{1}{m}\sum_{i=1}^{m}\bar{P}_i(L),
\label{eq.7.2}
\end{equation}
where $m$ is the number of users. Larger MAP index corresponds to better overall ranking accuracy. Precision is defined as the ratio of the number of recommended objects appeared in the testing set to the total number of recommended objects. Mathematically, for all users, the average value of Precision is defined as
\begin{equation}
P(L) = \frac{1}{m}\sum_{i=1}^{m}\frac{d_i(L)}{L},
\label{eq.7}
\end{equation}
Recall is defined as the ratio of the number of recommended objects appeared in user's recommendation list to the total number of objects in the test set. Mathematically, for all users, the average value of Recall is defined as
\begin{equation}
R(L) = \frac{1}{m}\sum_{i=1}^{m}\frac{d_i(L)}{D(i)},
\label{eq.8}
\end{equation}
Larger Precision and Recall mean higher accuracy of the recommendation.

Diversity is an important metric in evaluating the variety of objects that are recommended by personalized recommendation algorithms. As it is hard to obtain the external sources of the object similarity information, the diversity measures are usually based on the rating matrix. One of the commonly used diversity metrics is Inter-similarity, which can be quantified by Hamming distance \cite{Zhou2008}. The average value of Hamming distance for all users is defined as
\begin{equation}
H(L) = \frac{1}{m(m-1)}\sum_{i=1}^{m}\sum_{j=1}^{m}(1-\frac{C(i,j)}{L}),
\label{eq.9}
\end{equation}
where $C(i,j)=|o_i^L \cap o_j^L |$ is the number of common objects in user $i$'s and $j$'s recommendation lists. Larger value of Hamming distance corresponds to higher diversity. Another diversity metric is Intra-similarity \cite{ZhouT2009a}, which is measured by the cosine similarity between objects appeared in target user's recommendation list. Mathematically, the average value of Intra-similarity for all users is defined as
\begin{equation}
I(L) = \frac{1}{mL(L-1)}\sum_{i=1}^{m}\sum_{o_\alpha, o_\beta \in o_i^L, \alpha \neq \beta}S_{\alpha\beta}^{Cos},
\label{eq.10}
\end{equation}
where $S_{\alpha\beta}^{Cos}$ is the cosine similarity between objects $\alpha$ and $\beta$ in user $i$'s recommendation list $o_i^L$ with length $L$. Actually, the Intra-similarity index has been widely used in recommendation performance evaluation \cite{ZhouY2013,ChenW2013}. Smaller value of Intra-similarity means higher diversity of the recommendation.

Novelty \cite{Lu2012} is an important metric aiming to quantify the ability of an algorithm to generate novel (\emph{i.e.}, unpopular) and unexpected results. Here, we use the average Popularity of the recommended objects to quantify the novelty, which is defined as
\begin{equation}
N(L) = \frac{1}{mL} \sum_{i=1}^m{\sum_{o_\alpha\in o_i^L}k_{\alpha}},
\label{eq.11}
\end{equation}
where $k_{\alpha}$ is the degree of object $\alpha$ in user $i$'s recommendation list $o_i^L$. Smaller value of Popularity indicates higher novelty and potentially better user experience.

\section{Experiments and results}
\subsection{Performance of recommendation}
We apply the CosRA-based method to the four real online rating data sets. By comparison, some benchmark methods are also considered, including global ranking (GR), user-based collaborative filtering (UCF), item-based collaborative filtering (ICF), mass diffusion (MD) and heat conduction (HC). In GR, all objects are sorted in the descending order of their degrees and those with the largest degrees are recommended \cite{Zhou2007}. In UCF, the target user will be recommended objects collected by the users sharing similar tastes \cite{Liu2009}. Analogously, in ICF, the target user will be recommended objects similar to the ones that he preferred in the past \cite{ZhouT2009a}. We adopt the cosine similarity to quantify the user and object similarity in UCF and ICF, respectively. MD and HC both can be considered as resource-allocation processes on the user-object bipartite networks \cite{ZhouT2010,Ou2007}. Nevertheless, they have several distinguishing characteristics. The total amount of resources is conserved in MD instead of in HC. The transformation matrices in MD and HC are mutually transposed as the matrix is normalized by column in MD and by row in HC. Details of implementing the five benchmark methods can be found in the survey paper \cite{Lu2012}.

\begin{table}
\centering
\caption{Values of the six evaluation metrics after applying different recommendation algorithms on the four data sets. The length of recommendation list is set as $L=50$. The results are averaged over 10 independent realizations. For each data set and each evaluation metric, the best result is emphasized by bold.}
\begin{tabular*}{0.98\textwidth}{@{\extracolsep{\fill}}lccccccc}
\toprule
MovieLens-\scriptsize{100K} & AUC & MAP & $P$ & $R$ & $H$ & $I$ & $N$ \\
\midrule
GR & 0.863 & 0.208 & 0.058 & 0.358 & 0.395 & 0.408 & 255 \\
UCF & 0.887 & 0.315 & 0.070 & 0.476 & 0.550 & 0.394 & 242 \\
ICF & 0.888 & \textbf{0.385} & 0.073 & 0.494 & 0.674 & 0.413 & 211 \\
MD & 0.898 & 0.325 & 0.075 & 0.527 & 0.618 & 0.355 & 230 \\
HC & 0.842 & 0.037 & 0.021 & 0.123 & \textbf{0.858} & \textbf{0.056} & \textbf{23} \\
CosRA & \textbf{0.908} & 0.380 & \textbf{0.082} & \textbf{0.575} & 0.724 & 0.335 & 204 \\
\midrule
MovieLens-\scriptsize{1M} & AUC & MAP & $P$ & $R$ & $H$ & $I$ & $N$ \\
\midrule
GR & 0.856  & 0.144  & 0.053  & 0.222  & 0.403  & 0.415  & 1660 \\
UCF & 0.872  & 0.176  & 0.061  & 0.263  & 0.458  & 0.415  & 1640 \\
ICF & 0.885  & \textbf{0.289}  & 0.072  & 0.314  & 0.629  & 0.404  & 1445 \\
MD & 0.885  & 0.188  & 0.066  & 0.297  & 0.504  & 0.403  & 1618 \\
HC & 0.881  & 0.052  & 0.034  & 0.162  & \textbf{0.861}  & \textbf{0.045}  & \textbf{198} \\
CosRA & \textbf{0.895}  & 0.223  & \textbf{0.074}  & \textbf{0.350}  & 0.598  & 0.387  & 1541 \\
\midrule
Netflix & AUC & MAP & $P$ & $R$ & $H$ & $I$ & $N$ \\
\midrule
GR & 0.933  & 0.161  & 0.043  & 0.370  & 0.356  & 0.374  & 2416 \\
UCF & 0.939  & 0.196  & 0.047  & 0.411  & 0.406  & 0.375  & 2385 \\
ICF & 0.937  & \textbf{0.240}  & \textbf{0.051}  & 0.427  & 0.556  & 0.374  & 2065 \\
MD & 0.948  & 0.207  & 0.048  & 0.426  & 0.426  & 0.368  & 2369 \\
HC & 0.889  & 0.002  & 0.001  & 0.024  & \textbf{0.796}  & \textbf{0.004}  & \textbf{15} \\
CosRA & \textbf{0.950}  & 0.229  & \textbf{0.051}  & \textbf{0.449}  & 0.482  & 0.361  & 2298 \\
\midrule
RYM & AUC & MAP & $P$ & $R$ & $H$ & $I$ & $N$\\
\midrule
GR & 0.855  & 0.057  & 0.005  & 0.160  & 0.069  & 0.143  & 1245 \\
UCF & 0.919  & 0.175  & 0.015  & 0.417  & 0.759  & 0.167  & 1124 \\
ICF & 0.932  & \textbf{0.352}  & 0.017  & 0.445  & 0.914  & 0.177  & 656 \\
MD & 0.941  & 0.209  & 0.018  & 0.471  & 0.789  & 0.155  & 1089 \\
HC & 0.933  & 0.130  & 0.014  & 0.361  & \textbf{0.949}  & \textbf{0.057}  & \textbf{214} \\
CosRA & \textbf{0.952}  & 0.292  & \textbf{0.019}  & \textbf{0.482}  & 0.879  & 0.144  & 819 \\
\bottomrule
\end{tabular*}
\label{tab2:AUC}
\end{table}

Results of the seven evaluation metrics are shown in Table~\ref{tab2:AUC}. When focusing on the accuracy, CosRA-based method has the best performance on all the four data sets, as indicated by the highest values of AUC, Precision and Recall. The AUC values for the CosRA-based method are 0.908, 0.895, 0.950 and 0.952 for  MovieLens-\scriptsize{100K}\normalsize{, MovieLens-}\scriptsize{1M}\normalsize{,} Netflix and RYM, respectively. Although the CosRA-based method is slightly inferior to ICF as evaluated by the Mean Average Precision index, it has remarkable advantage towards the other four methods. GR and HC have poor performance as indicated by the generally smaller values of accuracy metrics, especially in Precision and Recall. When focusing on diversity, on the one hand, the values of Inter-similarity (Hamming distance) for the CosRA-based method are much larger than those in GR, UCF and MD and not far behind those in ICF and HC. On the other hand, the values of Intra-similarity for the CosRA-based method are smaller than those in GR, UCF and MD. These results suggest that the CosRA-based method has advantage in diversity although it is a little inferior to HC as evaluated by Inter-similarity and Intra-similarity. When focusing on novelty, the CosRA-based method remarkably outperforms GR, UCF and MD as indicated by the smaller values of Popularity, although ICF and HC perform best again. Indeed, it is challenging to solve the accuracy-diversity dilemma in recommender systems. Based on these observations, it can be concluded that the CosRA-based method has overall better accuracy, well diversity and novelty in personalized recommendation.

\subsection{Analysis of mechanisms}
\begin{figure}[t]
\centering
\includegraphics[width=0.65\textwidth]{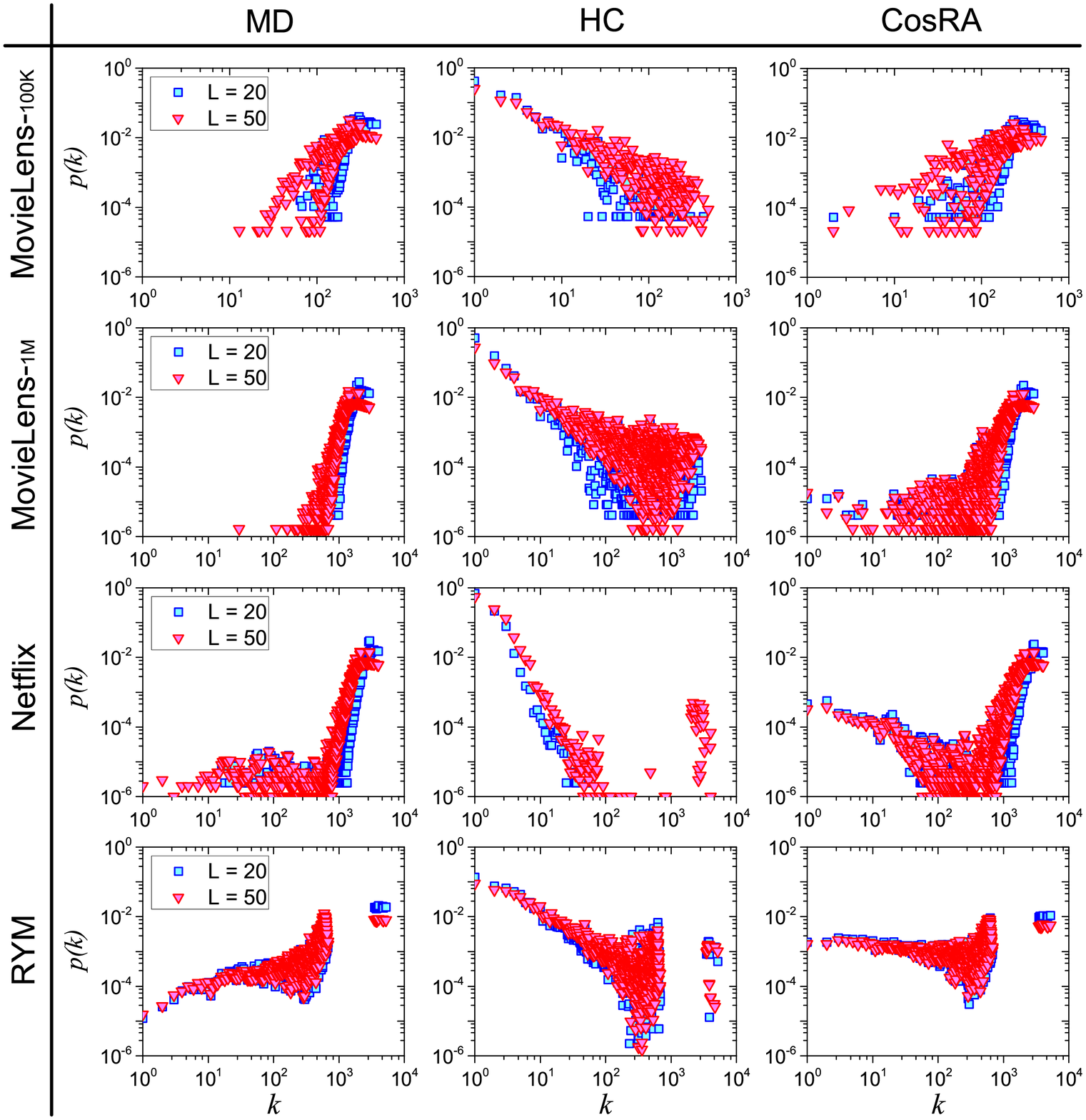}\\
\caption{Degree distribution of the recommended objects after applying MD, HC and CosRA-based methods on the four data sets. Results are shown for one realization in log-log plot. Blue squares and red triangles correspond to results under $L=20$ and $L=50$, respectively.}
\label{fig2:MHE}
\end{figure}

To better understand the mechanism of the CosRA-based method, we show the degree distributions of the recommended objects for all users in Fig.~\ref{fig2:MHE}. To make a comparison, MD and HC are also studied. In MD, there is a high probability for large-degree objects being recommended (see the first column of Fig.~\ref{fig2:MHE}), whereas HC prefers to recommend small-degree objects (see the second column of Fig.~\ref{fig2:MHE}). The two strong trends of MD and HC both have disadvantages, resulting in poor diversity and novelty of MD and low accuracy of HC. Fortunately, the CosRA-based method finds a balance among accuracy and diversity by recommending both large-degree and small-degree objects without any strong bias (see the last column of Fig.~\ref{fig2:MHE}).

\begin{figure}[t]
\centering
\includegraphics[width=\textwidth]{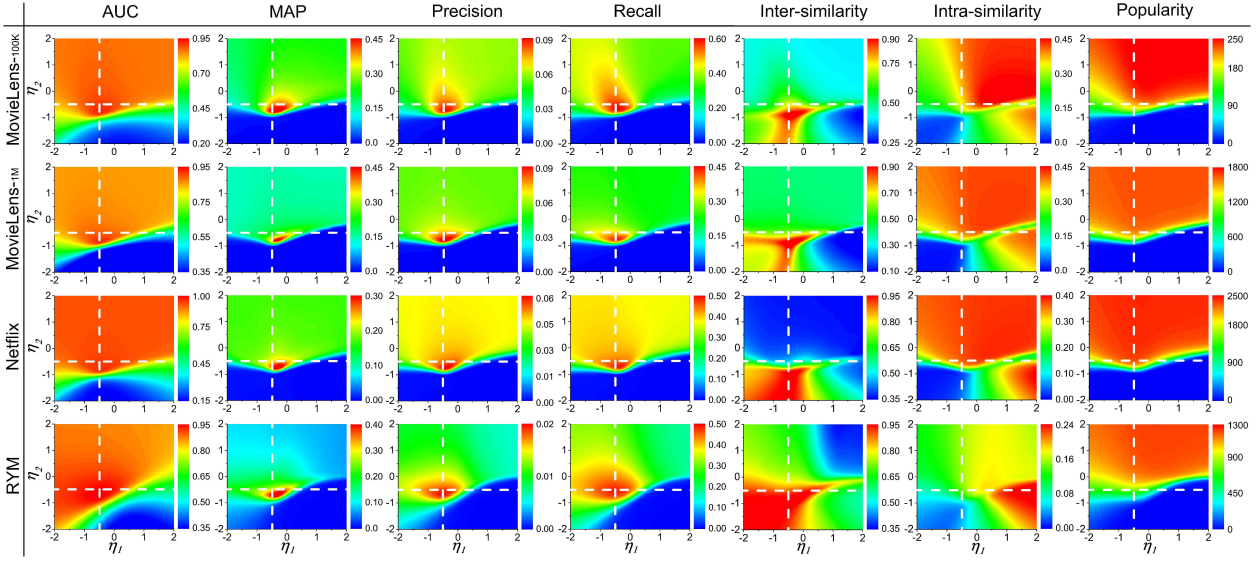}\\
\caption{Performance of the generalized CosRA-based method after tested on the four data sets. The parameters $\eta_{1}$ and $\eta_{2}$ are varying from $-2$ to $2$. Vertical and horizontal dash lines correspond to $\eta_{1}=-0.5$ and $\eta_{2}=-0.5$, respectively. The length of recommendation list is set as $L=50$ and the results are not sensitive to the value of $L$. The results are averaged over 10 independent realizations.}
\label{fig3:APRHIP}
\end{figure}

For a more systematic analysis on the CosRA index, we extend it to a more general form by introducing two turnable parameters, $\eta_{1}$ and $\eta_{2}$. Mathematically, the generalized CosRA index is formulated as
\begin{equation}
S_{\alpha\beta}^{CosRA*} = \frac{1}{(k_{\alpha} k_{\beta})^{-\eta_{2}}} \sum_{l=1}^{m} \frac{a_{l\alpha} a_{l\beta}}{(k_l)^{-2 \eta_{1}}}.
\label{eq:cosra2}
\end{equation}
Then the personalized recommendation algorithm based on the generalized CosRA index works as follows: Firstly, the resource of object $\alpha$ for user $i$ is initialized by Eq.~(\ref{eq:init}). Secondly, the resources of all objects are redistributed via the transformation $f'^{(i)}= S^{CosRA*} f^{(i)}$, where $f^{(i)}$ and $f'^{(i)}$ record all objects' initial and final resources, respectively. Finally, all objects are sorted by $f'^{(i)}$, and the top-$L$ uncollected objects are recommended to user $i$. Notice that the original CosRA index is a special case when $\eta_{1}=\eta_{2}=-0.5$. By varying $\eta_{1}$ and $\eta_{2}$, we study how the similarity index affects the performance of recommendation. As shown in Fig.~\ref{fig3:APRHIP}, the generalized CosRA-based method achieves its best performance when both $\eta_{1}$ and $\eta_{2}$ are around $-0.5$. Specifically, when focusing on accuracy, the values of AUC, MAP, Precision and Recall reach their maximum when $\eta_{1}$ and $\eta_{2}$ are around $-0.5$, as marked by vertical and horizontal dash lines in the first four columns of Fig.~\ref{fig3:APRHIP}. The accuracy metrics perform best at almost the same parameters on all data sets, which is a strong evidence that the optimal parameters, $\eta_{1}=-0.5$ and $\eta_{2}=-0.5$, for the generalized CosRA index are universal.

When focusing on diversity, the generalized CosRA-based method has better performance when $\eta_{1}$ and $\eta_{2}$ are smaller than $-0.5$, as indicated by the larger values of Inter-similarity (Hamming distance) and the smaller values of Intra-similarity in the fifth and sixth columns of Fig.~\ref{fig3:APRHIP}, respectively. When $\eta_{1}$ and $\eta_{2}$ exceed $-0.5$, the diversity of the generalized CosRA-based method largely decreases. When focusing on novelty, the diagrams are almost divided into two parts by $\eta_{2}\approx-0.5$ and the generalized CosRA-based method has remarkably lower Popularity (\emph{i.e.}, higher novelty) when $\eta_{2}<-0.5$ as shown in the last column of Fig.~\ref{fig3:APRHIP}. That is mainly because smaller $\eta_{2}$ benefits small-degree (\emph{i.e.}, unpopular) objects in receiving resources. After a comprehensive consideration, it can be concluded that the original parameters, $\eta_{1}=-0.5$ and $\eta_{2}=-0.5$, are almost optimal and the effectiveness of the generalized CosRA index cannot be remarkably improved by adjusting the two parameters.

\section{Conclusions and discussion}
In summary, we have proposed a vertex similarity index for better personalized recommendation, which combines advantages of both the cosine index and the resource-allocation index. Based on the proposed index, we further propose a personalized recommendation algorithm. Extensive experiments on real data sets suggest that the proposed algorithm has better accuracy and well diversity and novelty compared with some benchmark methods. To further understand how the similarity index works, we show the degree distributions of the recommended objects for all users. Results suggest that the proposed method does not have strong bias on objects' degrees compared with other benchmark methods. Indeed, the similarity index finds a balance among the three important evaluation metrics and improves the overall algorithmic performance. Further, we extend the similarity index to a more general form, however, results suggest that the original similarity index is almost optimal. That is to say, the similarity index is free of parameters, which is a significant advantage in real applications.

Our work highlights the importance of the vertex similarity index in personalized recommendation and suggests that the adoption of suitable similarity index can enhance the algorithmic performance. By applying the novel similarity index to the personalized recommendation, not only the accuracy is improved, but also the well diversity and novelty are achieved. In fact, the similarity-based recommendation algorithm is similar to the previous hybrid method in the case of the hybridization parameter of the transition matrix being equal to 0.5 \cite{ZhouT2010}. However, from a different perspective, we contribute to propose a new vertex similarity index instead of a straightforward hybrid recommendation algorithm. Nevertheless, the consistent with the existing hybrid recommendation algorithm verifies the rationality of the proposed parameter-free similarity index and justifies that the index is very simple but effective in recommendation.

Moreover, how to balance the accuracy, diversity and novelty in recommender systems is still an open issue \cite{Lu2012}. Although the heat conduction method performs best for the diversity and novelty metrics, it has strong bias on objects' degrees, leading to the poor performance on accuracy metrics. Nevertheless, our work provides a promising way to deal well with the three metrics by applying a suitable vertex similarity index. In addition to focusing on solving the accuracy-diversity dilemma, recent work has further investigated the stability of similarity measurements for bipartite networks by using the average ranking position to describe the stability of the recommendation results \cite{Liu2016}. It has been pointed out that by using stable similarity measurements the performance of recommendation can be largely improved. Therefore, the properties and the evaluation of the proposed similarity index remains further investigation.

Further more, pairwise vertex similarity is a fundamental index for many network functions and physical systems \cite{Newman2003,Wang2009}. That is to say, the proposed similarity index can find applications in solving many network-related problems, such as link predication \cite{Lu2011l,Lu2015t}, community detection \cite{Newman2004,Pany2010,Chen2012}, spreading activation \cite{Thiel2010}, network evolution \cite{Pap2013,Ma2007}, web searching \cite{Blondel2004,Zhang2015}, data clustering \cite{Sawa2003,Santos2015}, and gene ranking \cite{Zhu2012}. By contrast, it would be hard for the hybrid transition matrix to be applied to solve these problems. As future works, we could consider designing more suitable similarity indices for networks \cite{Tsourakakis2014,Chen2015} and introducing reputation systems into the personalized recommendation to improve its robustness in resisting spamming attacks \cite{Gao2015,Resnick2000}.

\section*{Acknowledgments}
The authors acknowledge Hai-Xing Dai for useful discussions. This work was partially supported by the National Natural Science Foundation of China (Grant Nos. 11222543 and 61433014). TZ acknowledges the Program for New Century Excellent Talents in University (Grant No. NCET-11-0070), and the Special Project of Sichuan Youth Science and Technology Innovation Research Team (Grant No. 2013TD0006).

\end{document}